\documentclass[aps,pra,twocolumn,amsmath,amssymb,reprint]{revtex4-1}

\usepackage{graphicx}
\usepackage{amssymb}
\usepackage{amsmath}
\usepackage{bm}
\usepackage{xcolor}
\usepackage{verbatim}
\usepackage{dcolumn}
\usepackage{multirow}
\usepackage{url}
\usepackage{hyperref}
\usepackage[hyphenbreaks]{breakurl}

\begin{document}

\title{Dual-species Bose-Einstein condensates of $^{23}$Na and $^{41}$K with tunable interactions}
\author{Jaeryeong Chang, Sungjun Lee, Yoonsoo Kim, Younghoon Lim, and Jee Woo Park}
\email{jeewoopark@postech.ac.kr}

\affiliation{Department of Physics, Pohang University of Science and Technology, Pohang 37673, Republic of Korea}

\begin{abstract}
We report the creation of dual-species Bose-Einstein condensates~(BECs) of $^{23}$Na and $^{41}$K. Favorable background scattering lengths enable efficient sympathetic cooling of $^{41}$K via forced evaporative cooling of $^{23}$Na in a plugged magnetic trap and an optical dipole trap. The $1/e$ lifetime of the thermal mixture in the stretched hyperfine state exceeds 5~s in the presence of background scattering. At the end of evaporation, we create dual BECs in the immiscible phase, with about $3\times10^5$~$^{23}$Na atoms surrounding $5\times10^4$~$^{41}$K atoms. To further enable the tuning of the interspecies interaction strength, we locate multiple Feshbach resonances at magnetic fields up to 100~G. The broadest $s$-wave resonance located at 73.4(3)~G features a favorable width of 1.8(2)~G. This work sets the stage for the creation of ultracold gases of strongly dipolar bosonic $^{23}$Na$^{41}$K molecules as well as the exploration of many-body physics in bosonic $^{23}$Na-$^{41}$K mixtures.
\end{abstract}

\maketitle

\section{INTRODUCTION}

Quantum mixtures of atoms with tunable interactions have enabled the controlled exploration of strongly correlated phenomena. Realized in atomic mixtures of different hyperfine states~\cite{myatt1997production,stenger1998spin,granade2002all}, isotopes~\cite{truscott2001observation,schreck2001quasipure,papp2008tunable}, or species~\cite{modugno2002two,hadzibabic2002two,Taglieber2008mixture,mccarron2011dual,lercher2011production,park2012quantum,wacker2015tunable,Wang2016mixture,schulze2018feshbach,warner2021overlapping} near a Feshbach resonance, they give access to paradigmatic problems in many-body physics such as the emergence of superfluidity in strongly interacting Fermi gases~\cite{regal2004fermionic,zwierlein2005vortices}, coupled excitations in binary superfluids~\cite{ferrier2014mixture}, Bose and Fermi polarons~\cite{schirotzek2009polaron,kohstall2012metastability,hu2016bose,jorgensen2016observation}, and beyond mean-field effects in self-bound quantum droplets~\cite{cabrera2018quantum,semeghini2018self}. 

For heteronuclear mixtures, a particularly intriguing application lies in the creation of ultracold gases of polar molecules. Produced through a two-step process that involves the association of loosely bound dimers and the subsequent transfer to the tightly bound ground state, these molecules feature large electric dipole moments that reach up to a few Debye~\cite{Moses2017molecules}. This gives rise to strong long-range and anisotropic dipole-dipole interactions. Quantum degenerate gases of polar molecules are thus expected to realize a host of exotic many-body phases of matter such as quantum crystals~\cite{buchler2007strongly} and topological superfluids~\cite{cooper2009stable,levinsen2011topological}, and can be employed as efficient qubits for the construction of molecular quantum computers~\cite{demille2002quantum,park2017second,Ni2018gate,gregory2021robust,bao2022dipolar}.

\begin{figure}[b]
 \includegraphics[width=86mm]{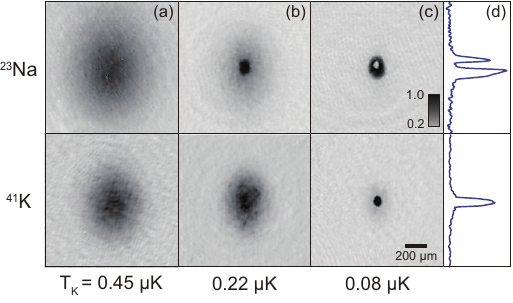}
  \caption{Absorption images of $^{23}$Na and $^{41}$K during evaporative cooling. (a), (b), and (c) are vertical sets of 10~ms time-of-flight (TOF) absorption images of $^{23}$Na and $^{41}$K. Temperatures are estimated from the thermal wings of $^{41}$K. Atom numbers of $^{23}$Na ($^{41}$K) are (a) $19~(2.8)\times10^5$, (b) $5.3~(2.5)\times10^5$ and (c) $2.4~(0.4)\times10^5$ for the shown images. (d) is the hole-sliced column density of (c). Phase separation of the mixture emerges at the trap center, due to the oblate trap geometry and the strong interspecies repulsion.}
\label{dual BECs}
\end{figure}

Since the first creation of fermionic $^{40}$K$^{87}$Rb~\cite{ni2008high}, ultracold polar molecules of diverse combinations of alkali atoms have been realized~\cite{takekoshi2014ultracold,molony2014creation,park2015ultracold,guo2016creation,rvachov2017long,voges2020ultracold,cairncross2021assembly,stevenson2023ultracold,he2023efficient}. Among them, NaK features a unique combination of strengths. First of all, thanks to the existence of stable isotopes of K, ultracold gases of both bosonic $^{23}$Na$^{39, 41}$K and fermionic $^{23}$Na$^{40}$K can be realized, even within a single experimental system. Not only does this give access to the study of strongly dipolar many-body systems in both Bose and Fermi gases, but also opens the door to highly consistent studies of the role of quantum statistics in chemical reactions~\cite{Ospelkaus2010chemical,chen2023ultracold,liu2023quantum}. 
Furthermore, NaK has a large electric dipole moment of 2.72~Debye~\cite{Gerdes2011NaK}, and its stability against binary exchange reactions in the absolute ground state can suppress two-body collisional losses~\cite{Zuchowski2010reaction,park2015ultracold}, which can be important, for example, in an optical lattice~\cite{Moses2015lattice}.

Experiments with ultracold NaK have achieved remarkable progress in recent years. Since the first creation of ultracold $^{23}$Na-$^{40}$K Bose-Fermi mixtures~\cite{park2012quantum}, ground state $^{23}$Na$^{40}$K molecules have been realized~\cite{park2015ultracold}, and their degenerate Fermi gases have been created via direct evaporative cooling~\cite{schindewolf2022evaporation}. For bosonic $^{23}$Na$^{39}$K, degenerate $^{23}$Na-$^{39}$K Bose-Bose mixtures~\cite{schulze2018feshbach} and ultracold ground state $^{23}$Na$^{39}$K molecules~\cite{voges2020ultracold} have been created. However, due to the large and negative background interspecies scattering length, elaborate interaction tuning is required for sympathetic cooling, which makes the realization of deeply degenerate $^{23}$Na-$^{39}$K mixtures challenging.

Here, we report the creation of dual-species Bose-Einstein condensates (BECs) of $^{23}$Na-$^{41}$K. Thanks to the favorable inter- and intraspecies background scattering lengths, we achieve highly efficient and simple sympathetic cooling of $^{41}$K with $^{23}$Na. At the end of evaporation, we create essentially pure BECs in their immiscible phase, with $3\times10^5$~$^{23}$Na atoms surrounding $5\times10^4$~$^{41}$K atoms. The lifetime of the thermal mixture in the presence of background collisions exceeds 15~s in the ground hyperfine state and 5~s even in the excited stretched state, facilitating the creation of large degenerate samples. In a highly oblate trap, the $^{41}$K BEC leaves a sharp density depletion near the center of the $^{23}$Na BEC, which forms a toroidal structure. To further enable the tuning of the interspecies scattering length, we locate seven Feshbach resonances in the $s$- and $p$-wave collision channels. The broadest $s$-wave resonance in the lowest hyperfine state combination features a width of 1.8(2)~G at 73.4(3) G, which enables precise control of the interaction strength between $^{23}$Na-$^{41}$K. These results serve as the stepping stone for the creation of strongly dipolar $^{23}$Na$^{41}$K bosonic ground state molecules, as well as the exploration of coupled superfluid dynamics and beyond mean-field effects in $^{23}$Na-$^{41}$K mixtures.

\section{DUAL-SPECIES MOT LOADING}

Our experiment employs a spin-flip Zeeman slower and a 2D$^{+}$ MOT to load $^{23}$Na and $^{41}$K atoms into the main vacuum chamber, respectively. The 2D$^{+}$ MOT glass cell contains about 30~mg of enriched potassium, which consists of 25.6\% $^{41}$K, 5.5\% $^{40}$K, and 68.9\% $^{39}$K. Note that the natural abundance of $^{41}$K is 6.7\%. By a simple manipulation of the laser cooling system, the 2D$^{+}$ MOT can generate atomic fluxes of either bosonic $^{41}$K or fermionic $^{40}$K~\cite{KPS}. This work focuses on $^{41}$K, where about $10^7$ atoms are loaded into the MOT in 5~s. For $^{23}$Na, a dark spot MOT is used to achieve high atomic densities~\cite{ketterle1993high}, and about $3\times10^9$ atoms are loaded into the MOT in 5~s.

When capturing the atoms in the MOT, we first load the $^{23}$Na atoms, perform molasses cooling and optical pumping to the magnetically trappable stretched $\vert$$F,m_{F}$$\rangle$ = $\vert$2,2$\rangle$ state, and then capture them in a magnetic trap. Here, $F$ is the total angular momentum quantum number, and $m_{F}$ is its projection. Then, we load the $^{41}$K atoms, while $^{23}$Na remains trapped in the magnetic field gradient of the $^{41}$K MOT. The separation of the MOT loading sequence prevents light-assisted collisional losses between $^{23}$Na-$^{41}$K, and it further allows us to independently optimize the field gradients of each MOT. Specifically, we observe that applying different magnetic field gradients of 6.5 and 9.7~G/cm for $^{23}$Na and $^{41}$K, respectively, optimizes the MOT performance. Once the $^{41}$K atoms are loaded into the MOT, we apply a compressed MOT phase and optical pumping before transferring the mixture to an optically plugged magnetic trap.

\begin{figure}[t]
 \includegraphics[width=0.48\textwidth]{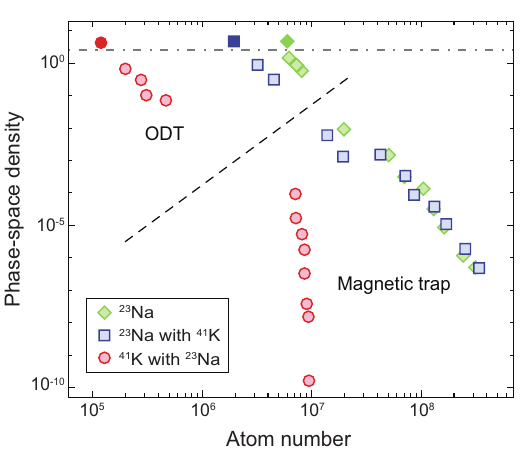}
  \caption{Phase-space density evolution of $^{23}$Na and $^{41}$K as a function of the atom number. Green diamonds: $^{23}$Na alone; Blue squares: $^{23}$Na with $^{41}$K; Red circles: $^{41}$K. Data points below the dashed line are measured in the plugged magnetic trap and those above are measured in the optical dipole trap. The gray dot-dashed line indicates the phase-space density above which Bose-Einstein condensation occurs.}
\label{PSD curve}
\end{figure}

\section{SYMPATHETIC COOLING}

\begin{figure*}[t]
\includegraphics[width=1\textwidth]{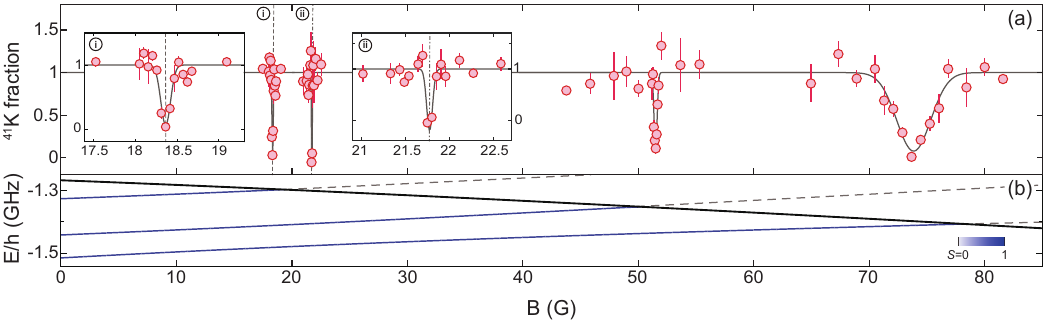}
  \caption{Feshbach loss spectroscopy of $^{23}$Na$\vert$1,1$\rangle$+$^{41}$K$\vert$1,1$\rangle$. (a) The fraction of $^{41}$K atoms that remain in the dipole trap after a variable hold time is shown as a function of the magnetic field. The hold times are optimized to maximize the visibility of each loss feature. Insets show zooms of the narrow loss features at 18.36(1)~G and 21.77(1)~G. Error bars are the standard error of the mean. (b) The bound state energies calculated in the asymptotic bound state model (blue solid lines) are shown together with the open-channel threshold (black solid line).
  The bound state color represents its spin character, from light (singlet) to dark (triplet) blue. At low magnetic fields, the resonances arise from dominantly triplet bound states.}
\label{FB11}
\end{figure*}

In the magnetic trap, we perform forced radio-frequency~(rf) evaporative cooling of $^{23}$Na on the $\vert2,2\rangle \rightarrow \vert$1,1$\rangle$ transition. $^{41}$K is sympathetically cooled via elastic collisions with $^{23}$Na. To protect the atoms from Majorana losses, we employ a blue-detuned plug laser beam (wavelength 532~nm, power 15~W, and waist 80~$\mu$m), which is aligned close to the magnetic trap center. During rf-evaporation, the magnetic field gradient $B'$ is kept at 80~G/cm for the first 8~s to maintain a high elastic collision rate. Then, $B'$ is monotonically decompressed to 26~G/cm within the last 10~s to reduce inelastic losses. The frequency of the rf source driving the $^{23}$Na hyperfine transition is decreased from 1840 to 1773.5~MHz during this time. At the end of rf-evaporation, the mixture, which has a temperature of 6.7(1)~$\mu$K, is transferred to an oblate optical dipole trap (ODT). 

Our single beam ODT (wavelength 1064~nm and maximum power 9~W) is tightly focused along the gravitational direction with a waist of $w_z = 12~\mu$m and transverse waist of $w_x=615~\mu$m. The initial dipole trap depth is $U_{\text{Na}}=$ $k_{\text{B}}~\times$ 40~$\mu$K for $^{23}$Na and $U_{\text{K}}=$ $k_{\text{B}}~\times$ 102~$\mu$K for $^{41}$K, where $k_{\text{B}}$ is the Boltzmann constant. In the ODT, the mixture is further evaporatively cooled by reducing the trap depth. Since the trap depth for $^{41}$K is approximately twice that of $^{23}$Na, $^{23}$Na atoms are dominantly removed from the trap.

At the end of evaporation, we create dual BECs of $^{23}$Na and $^{41}$K, as shown in Fig.~\ref{dual BECs}. For the coldest samples created, both BECs are essentially pure, and each BEC contains $N_{\text{Na}}=2.9(2)\times10^5$ and $N_{\text{K}}=4.9(1)\times10^4$ atoms. At the final ODT trap depth $U_{\text{Na}}=$ $k_{\text{B}}~\times$ 1.7~$\mu$K for $^{23}$Na and $U_{\text{K}}=$ $k_{\text{B}}~\times$ 4.4~$\mu$K for $^{41}$K, the trapping frequencies are measured to be $[\omega^{\textrm{Na}}_{x},\omega^{\textrm{Na}}_{y},\omega^{\textrm{Na}}_{z}]=$~$2\pi$~$\times~[9(1), 5(1), 476(11)]$~Hz and $\omega^{\text{K}}~\simeq~1.12\times\omega^{\text{Na}}$. The approximate Thomas-Fermi (TF) radii are $[R^{\textrm{Na}}_x, R^{\textrm{Na}}_y, R^{\textrm{Na}}_z] = [69(1), 130(2), 1.4(1)]$$~\mu$m and $[R^{\textrm{K}}_x, R^{\textrm{K}}_y, R^{\textrm{K}}_z] = [30(1), 57(2), 0.6(1)]$$~\mu$m. Here, $R^{\textrm{Na}}_{x,y}$ are obtained from fits neglecting the density-depleted hole in the sample, and $R_z$ is estimated from trap parameters for both species. Our highly oblate trap geometry gives rise to a differential gravitational sag of $\Delta z$ $\simeq 0.33~\mu$m, which, compared to the TF radii of the samples, ensures good spatial overlap between the atomic clouds along the $z$-axis.

Due to the repulsive interaction in their stretched hyperfine states, the two BECs are spatially separated. The background scattering lengths of $^{23}$Na and $^{41}$K are 64.30(40)~$a_0$ and 60.54(6)~$a_0$~\cite{Knoop2011NaFB,Falke200841K}, respectively, and between $^{23}$Na-$^{41}$K, it is predicted to be 267~$a_0$~\cite{viel2016feshbach}. For our number-imbalanced samples, the $^{23}$Na atoms condense first at the trap center (Fig.~\ref{dual BECs}(b)). This implies that $^{41}$K should start to condense at the boundary of the $^{23}$Na BEC to minimize the repulsive energy cost. However, at the end of evaporation, we observe that the $^{41}$K BEC exists near the trap center, and it fully depletes the $^{23}$Na BEC along the tightly confining direction (Fig.~\ref{dual BECs}(c)). This non-trivial formation dynamics of the dual BECs in a highly oblate trap will be further explored in future works.
 
To characterize the efficiency of sympathetic cooling, we measure the phase-space density (PSD) evolution of the mixture, as shown in Fig.~\ref{PSD curve}. Here, temperatures are extracted from the thermal wings of the clouds after 10~ms time-of-flight (TOF), and atom numbers are extracted from the images assuming resonant absorption cross-sections. The steep increase of the $^{41}$K PSD in the magnetic trap during rf-evaporation shows that $^{23}$Na is an efficient sympathetic coolant for $^{41}$K. The cooling efficiency $\Gamma$ = -$d$ln(PSD)/$d$ln($N$), where $N$ is the atom number, reaches $\Gamma_{\text{K}} = 35$ for $^{41}$K, which is remarkably higher than that of other mixtures~\cite{wu2011strongly,park2012quantum,warner2021overlapping}. $\Gamma_{\text{Na}}$ for $^{23}$Na is given by 3.7, and in the presence of $^{41}$K, it is slightly reduced to 3.0 due to the added thermal load. In the dipole trap, sympathetic cooling becomes less efficient with $\Gamma_{\text{Na}} = 3.1$ and $\Gamma_{\text{K}} = 2.7$. This reduction is most likely from the enhanced collisional losses at higher atomic densities.

\section{FESHBACH RESONANCES}

Tuning the interaction strength using a Feshbach resonance is essential for the exploration of many-body physics with ultracold atoms and also for the efficient formation of Feshbach molecules. For Na-K, Feshbach resonances in $^{23}$Na-$^{40}$K Bose-Fermi mixtures and $^{23}$Na-$^{39}$K Bose-Bose mixtures have been previously investigated~\cite{park2012quantum,schulze2018feshbach}. These works have led to the refinement of Na-K ground state potentials via coupled-channel calculations~\cite{viel2016feshbach, hartmann2019feshbach}. Based on these theoretical investigations and the asymptotic bound state model (ABM) calculations~\cite{tiecke2010broad} that we have performed, we proceed with atom-loss spectroscopy to locate $^{23}$Na-$^{41}$K Feshbach resonances. To this end, we initially prepare a thermal mixture of $^{23}$Na-$^{41}$K in the dipole trap with about 5$-$10 $\times$ $10^4$ atoms per species in their stretched $\vert$2,2$\rangle$ states. Then, we apply a sequence of rf Landau-Zener sweeps to transfer the atoms to the desired hyperfine states. In this work, we search for resonances in three different combinations of $^{23}$Na$\vert$1,1$\rangle$+$^{41}$K$\vert$1,$\textit{m}_\textit{F}$$\rangle$, with spin states $\textit{m}_\textit{F}$ = 1, 0, and $-1$. To access the enhanced inelastic loss features near a Feshbach resonance, we ramp the magnetic field to a targeted value in 40~ms, apply a variable hold time of 0.25$-$6~s, and then ramp the field down to 2~G. Finally, absorption images are recorded, from which we extract the atom numbers. Fig.~\ref{FB11}(a) shows a representative loss spectrum in $^{23}$Na$\vert$1,1$\rangle$+$^{41}$K$\vert$1,1$\rangle$. 

\begin{table}[t]
\caption{\label{tab:table1}
 Summary of the interspecies Feshbach resonances between $^{23}$Na and $^{41}$K. The resonance positions and widths are determined by applying phenomenological Gaussian fits ($e^{-(B-B_\textrm{exp})^{2}/\Delta _\textrm{exp}^2}$) to the loss features. For the features at 18.36~G and 21.77~G, theoretical calculations predict the existence of both $s$- and $p$-wave resonances near this field. Hence, we do not attempt to further assign their resonance types.}
\begin{ruledtabular}
\begin{tabular}{cccc}
 \multirow{2}{*}{Entrance channel} & \multirow{2}{*}{$B_\textrm{exp}~\textrm{(G)}$} & \multirow{2}{*}{$\Delta_\textrm{exp}~\textrm{(G)}$} & Resonance\\ & & & type\\
\hline 
\rule{0pt}{2.5ex}$^{23}$Na$\vert$1,1$\rangle$+$^{41}$K$\vert$1,1$\rangle$
& 18.36(1)  & 0.08(1)  & $s$ or $p$\\
& 21.77(1)  & 0.06(1) & $s$ or $p$\\
& 51.55(1) & 0.18(2) & $s$\\
& 73.4(3)  & 1.8(2) & $s$\\
$^{23}$Na$\vert$1,1$\rangle$+$^{41}$K$\vert$1,0$\rangle$ 
& 67.39(8) & 0.32(9) &$s$\\
& 87.1(4) &0.9(3) &$s$\\
$^{23}$Na$\vert$1,1$\rangle$+$^{41}$K$\vert$1,$-1$$\rangle$ 
 & 106.4(3) & 1.2(5) & $s$
\end{tabular}
\end{ruledtabular}
\label{FBtotal}
\end{table}

From our search, we locate seven interspecies Feshbach resonances in the $s$- and $p$-wave collision channels, as summarized in Table~\ref{tab:table1}. Particularly, we find an $s$-wave resonance at 73.4(3)~G with a width of 1.8(2)~G in $^{23}$Na$\vert$1,1$\rangle$+$^{41}$K$\vert$1,1$\rangle$. The relatively broad width compared to its position makes this resonance appealing for precision control of the interspecies scattering length and for the creation of Feshbach molecules. A simple ABM calculation that neglects the open channel coupling shows that the $s$-wave resonances found in this work arise from dominantly triplet bound states. Since the singlet states only weakly mix with the triplet states in the explored magnetic field regime, we fix the singlet bound state energy to a value extracted from the NaK ground-state potential ($E_{\text{s}} \approx 806$~MHz)~\cite{hartmann2019feshbach} and perform a least-squares fit to optimize the triplet bound state energy ($E_{\text{t}} \approx 2085$~MHz). The calculated resonance pattern matches well with the observed experimental data (see Fig. 3(b)). The resonance positions and widths are also in good agreement with the coupled-channel predictions from Ref.~\cite{viel2016feshbach}.

 \begin{figure}[t]
    \includegraphics[width=86mm]{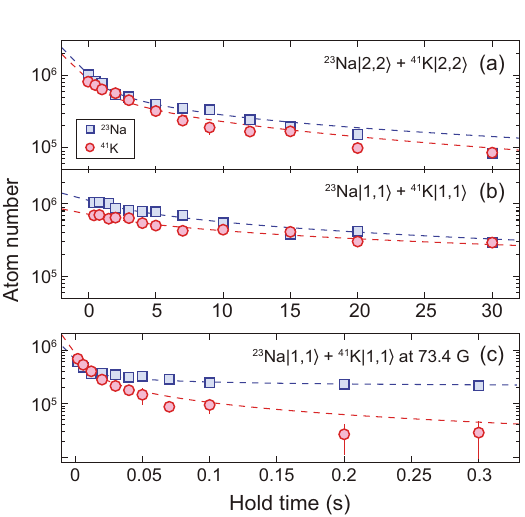}
    \caption{Lifetimes of the $^{23}$Na-$^{41}$K thermal mixture in the dipole trap. (a) $^{23}$Na$\vert$2,2$\rangle$+$^{41}$K$\vert$2,2$\rangle$ at 2~G, (b) $^{23}$Na$\vert$1,1$\rangle$+$^{41}$K$\vert$1,1$\rangle$ at 2~G, and (c) $^{23}$Na$\vert$1,1$\rangle$+$^{41}$K$\vert$1,1$\rangle$ at unitarity. Blues squares (red circles) indicate the atom number of $^{23}$Na ($^{41}$K) and error bars are the standard error of the mean. Dotted lines in (a) and (b) are guides to the eye, and lines in (c) indicate the fit of coupled differential equations governing the loss dynamics of the mixtures.}
    \label{Lifetime}
\end{figure}

\section{LIFETIMES}

To characterize the collisional stability of the mixture, we measure its lifetime in the presence of background collisions, as shown in Fig.~\ref{Lifetime}. For this, we prepare thermal clouds of $^{23}$Na-$^{41}$K in their stretched $\vert$2,2$\rangle$ state (Fig.~\ref{Lifetime}(a)) or ground $\vert$1,1$\rangle$ state (Fig.~\ref{Lifetime}(b)) in the dipole trap at a magnetic field of 2~G, away from any scattering resonances. The atom numbers are set close to $10^6$ for both species, and the temperature of the mixture is at 0.53(1)~$\mu$K, which is slightly above the critical temperature for $^{41}$K. The lifetime is measured by applying a variable hold time and extracting the number of remaining atoms. 

For both states, we observe an early-time non-exponential decay, which is most likely due to the combined effects of interspecies three-body losses that cause evaporative heating and single particle losses from evaporative cooling. Afterward, only single-body losses persist. Note that the triplet (singlet) background scattering length between $^{23}$Na and $^{41}$K is predicted to be 267 (-3.65)~$a_0$~\cite{viel2016feshbach}. For the $\vert$2,2$\rangle$ state combination, the mixture additionally experiences inelastic dipolar losses. The measured 1/$e$ lifetimes in $^{23}$Na$\vert$2,2$\rangle$+$^{41}$K$\vert$2,2$\rangle$ are 6.0(5) and 5.8(3)~s for $^{23}$Na and $^{41}$K, respectively, while in $^{23}$Na$\vert$1,1$\rangle$+$^{41}$K$\vert$1,1$\rangle$, are 16(1) and 20(2)~s. These can be compared to the lifetime observed in $^{23}$Na-$^{39}$K mixtures, which was about 240~ms due to the large and attractive background interspecies interaction~\cite{schulze2018feshbach}. 

We also explore the lifetime of the mixture at the broadest $s$-wave resonance (see Fig.~\ref{Lifetime}(c)). Here, after preparing the thermal mixture in the $^{23}$Na$\vert$1,1$\rangle$+$^{41}$K$\vert$1,1$\rangle$ state, the magnetic field is rapidly ramped across the low-lying Feshbach resonances in 50~ms and then jumped to unitarity at 73.4~G in 5~ms before applying a variable hold time. In this case, we observe a rapid decay of the atom number of both species with characteristic time scales of $\sim$100~ms (Fig.~\ref{Lifetime}(c)). The loss rate is significantly higher in $^{41}$K, which demonstrates that the loss mechanism at unitarity is dominated by the three-body Na-K-K collisions~\cite{Mikkelsen2015threebody}. 
To quantify the three-body loss coefficients, we fit the observed atom losses to a coupled loss model describing the three-body loss dynamics of the mixture~\cite{Wacker2016hetero}:

 \begin{align*}
    \frac{dN_{\mathrm{Na}}}{dt} {=} -\frac{2}{3}L_{\text{NNK}}\int n_{\mathrm{Na}}^{2}n_{\mathrm{K}}d^{3}r-\frac{1}{3}L_{\text{NKK}}\int n_{\mathrm{Na}}n_{\mathrm{K}}^{2}d^{3}r \\
     \frac{dN_{\mathrm{K}}}{dt} {=} -\frac{1}{3}L_{\text{NNK}}\int n_{\mathrm{Na}}^{2}n_{\mathrm{K}}d^{3}r-\frac{2}{3}L_{\text{NKK}}\int n_{\mathrm{Na}}n_{\mathrm{K}}^{2}d^{3}r.
\end{align*}

Here, $N_{\mathrm{Na(K)}}$ and $n_{\mathrm{Na(K)}}$ are the atom numbers and the atomic densities for each species, and $L_{\mathrm{NNK(NKK)}}$ is the three-body loss coefficient for the Na-Na-K (Na-K-K) channel. From the fits, we obtain $L_{\mathrm{NNK}}=7.4(3)\times10^{-25}~\textrm{cm}^{6}\textrm{s}^{-1}$ and $L_{\mathrm{NKK}}=4.3(1)\times10^{-23}~\textrm{cm}^{6}\textrm{s}^{-1}$. Since the loss rate in this regime is inversely proportional to the temperature squared~\cite{Rem2013threebody}, the scaled loss rate $\lambda_3=L_\textrm{NKK}T^2$ is extracted to be 1.2(1)$\times10^{-23}~\textrm{cm}^{6}\mu\textrm{K}^2\textrm{s}^{-1}$ for the dominant loss channel. This value is comparable to the loss rate observed in unitary K-Rb Bose-Bose mixtures~\cite{Wacker2016hetero} but different from homonuclear mixtures with $\lambda_3=3.0(3)\times10^{-20}~\textrm{cm}^{6}\mu\textrm{K}^2\textrm{s}^{-1}$~\cite{Eismann2016threebody}. Note that $\lambda_{3}$ is not expected to be universal for heteronuclear mixtures~\cite{Petrov2015threebody}.

\section{SUMMARY AND OUTLOOK}

In conclusion, we have created a degenerate Bose-Bose mixture of $^{23}$Na-$^{41}$K and revealed their interspecies Feshbach resonances. The long mixture lifetimes in the presence of background collisions dramatically simplify the sympathetic cooling process, since it does not require the mixture to be prepared near any Feshbach resonances. In fact, fine-tuning the interspecies scattering length near a Feshbach resonance during evaporative cooling can further enhance the cooling efficiency in the dipole trap. This will lead to the creation of even larger dual BECs. For the prospect of creating quantum degenerate gases of polar molecules, preparing larger atomic samples near quantum degeneracy will increase the molecule number that can be evaporatively cooled in the presence of DC or AC electric fields~\cite{Valtolina2020KRb,schindewolf2022evaporation, Bigagli2023MWevap,Lin2023MWevap}. 

Furthermore, the highly oblate $^{23}$Na-$^{41}$K dual BECs produced in this work give a unique opportunity to explore unusual non-equilibrium dynamics in coupled superfluids. For example, understanding the formation of dual BECs in an oblate trap and its connection to the underlying interactions may open up a new route to realize ball-shell structured BECs~\cite{zobay2001shell,Carollo2022NASA,Jia2022shell}, with its distinctive collective modes~\cite{Lannert2007collec,padavic2017collec,Sun2018collec}, thermodynamics~\cite{Totoni2020ther}, and vortex dynamics in a closed-surface topology~\cite{padavic2020vortex}. Also, the exploration of beyond mean-field effects in quantum droplets, particularly the role of quantum fluctuations in the 3D-2D crossover, is within reach~\cite{Zin2018droplet}.

\section*{ACKNOWLEDGEMENTS}

We thank Seokmin Jang for experimental assistance and Sooshin Kim for the critical reading of the manuscript. This work was supported by the National Research Foundation of Korea (Grants No. NRF-2020R1C1C1011092, NRF-2021R1A6A1A10042944), the Samsung Science and Technology Foundation (Project No. SSTF-BA2001-06), and the Ministry of Science and ICT, Korea, under
the ITRC support program (Grant No. IITP-2022-RS-2022-00164799) supervised by the IITP. Y. Lim acknowledges the support from the NRF BK21 FOUR Postdoctoral Fellowship and S. Lee from the Hyundai Motor Chung Mong-Koo Foundation. 

\bibliographystyle{apsrev4-1}

\end{document}